\documentclass[12pt, fleqn]{article}
\usepackage[cp1251]{inputenc}
\usepackage{latexsym,amsfonts,amssymb}
\usepackage{graphicx}

\usepackage{amsbsy}
\usepackage{amsmath}
\usepackage{epsf}

\newtheorem{theo}{Theorem}
\newtheorem{definition}{Definition}

\newcommand{\bt}{\begin{theo}}
\newcommand{\et}{\end{theo}}
\newcommand{\bd}{\begin{displaymath}}
\newcommand{\ed}{\end{displaymath}}

\newcommand{\lf}{\left}
\newcommand{\rg}{\right}

\newcommand{\be} {\begin{equation}}
\newcommand{\ee} {\end{equation}}
\newcommand{\ba}{\begin{array}{l}}
\newcommand{\ea} {\end{array}}
\newcommand{\bea}{\begin{eqnarray}}
\newcommand{\eea} {\end{eqnarray}}

\newcommand{\p} {\partial}

\begin{document}

\begin{center}
 {\Large \textbf{ Comments on the paper `Modelling and nonclassical symmetry analysis of a complex porous media
flow in a dilating channel'} }
\medskip\\
{\bf Roman Cherniha~$^{\dag,\dag\dag}$ \footnote{\small
Corresponding author. E-mails: r.m.cherniha@gmail.com;
roman.cherniha1@nottingham.ac.uk}}
 \\
{\it $^{\dag}$~National University of Kyiv-Mohyla Academy,\\
2, Skovoroda Street, Kyiv  04070, Ukraine\\
  $^{\dag\dag}$~School of Mathematical Sciences, University of Nottingham,\\
  University Park, Nottingham NG7 2RD, UK
}
 \end{center}

\begin{abstract}
The Comments are   devoted to the  recently published   paper  'Modelling and nonclassical symmetry analysis of a complex porous media
flow in a dilating channel' (Physica D.
481 (2025) 134834), in which a model describing an unsteady two-dimensional viscous incompressible fluid flow through a porous medium is studied. The main theoretical results of that study consists of finding Lie and nonclassical symmetries of a fourth-order PDE, which was derived by simplification of the given model.
Here it   is shown that the main theoretical results derived therein are incomplete and misleading.

\end{abstract}


The recent paper \cite{mandal-2025}  is   devoted to study a
mathematical  model describing an unsteady two-dimensional viscous,
incompressible fluid flow through a porous medium. The model
consists of the three-component nonlinear system (1)--(3) (see
\cite{mandal-2025}) and corresponding boundary conditions. It should
be stressed that this system with  $\nu=0$, i.e. without kinematic
viscosity, is nothing else but the famous Navier-Stokes system in 2D
space, what is, surprisingly, not indicated in that paper. Using
scaling transformations and introducing the stream function
$\Psi(x,y,t)$, the authors reduce the three-component system to the
single fourth-order PDE \cite{mandal-2025} \be \label {1}\ba
\frac{Re}{\epsilon}\Big( \frac{\partial^{3}\Psi}{\partial t\partial
y^{2}}+\frac{\partial^{3}\Psi}{\partial t\partial x^{2}}\Big)+
\frac{Re}{\epsilon^2}\Big( \frac{\partial \Psi}{\partial y}
\frac{\partial^{3}\Psi}{\partial x\partial y^{2}}-\frac{\partial
\Psi}{\partial x} \frac{\partial^{3}\Psi}{\partial y^{3}} \Big)
-\frac{Re}{\epsilon^2}\Big( \frac{\partial \Psi}{\partial x}
\frac{\partial^{3}\Psi}{\partial y\partial x^{2}}-\frac{\partial
\Psi}{\partial y} \frac{\partial^{3}\Psi}{\partial x^{3}} \Big)=\\
-\frac{1}{D_A}\Big( \frac{\partial^{2}\Psi}{\partial x^{2}}+
\frac{\partial^{2}\Psi}{\partial y^{2}} \Big)+\lambda
\Big(\frac{\partial^{4}\Psi}{\partial
x^{4}}+\frac{\partial^{4}\Psi}{\partial
y^{4}} +2\frac{\partial^{4}\Psi}{\partial x^2\partial y^{2}}\Big)\\
+\frac{\alpha}{\epsilon}\Big( 2(\frac{\partial^{2}\Psi}{\partial
x^{2}}+ \frac{\partial^{2}\Psi}{\partial
y^{2}})+x(\frac{\partial^{3}\Psi}{\partial
x^{3}}+\frac{\partial^{3}\Psi}{\partial x \p
y^{2}})+y(\frac{\partial^{3}\Psi}{\partial
y^{3}}+\frac{\partial^{3}\Psi}{\partial y \p
x^{2}})+2t(\frac{\partial^{3}\Psi}{\p t\partial
x^{2}}+\frac{\partial^{3}\Psi}{\partial t \p y^{2}}) \Big), \ea \ee
where all coefficients are some positive constants. In Theorem 1
\cite{mandal-2025}, the authors claim that Eq.(\ref{1}) admits an
infinite-dimensional Lie algebra generated by the Lie symmetries
(21)\cite{mandal-2025}. However, the authors missed the special case
$\epsilon=2\alpha D_A$, in which Eq.(1) admits another
infinite-dimensional Lie algebra. This algebra is generated by the
infinitesimal generators
\[\ba  X_1=(Re-2\alpha t)\frac{\p}{\p t}, \ X_2=F_1(t)\frac{\p}{\p\Psi},
\ X_3= y\frac{\p}{\p x}-x\frac{\p}{\p y}, \\ X_4= F_2(t)\frac{\p}{\p
x}+\frac{\epsilon\,y}{Re}\Big(\alpha F_2(t)+(Re-2\alpha
t)F_2'(t)\Big)\frac{\p}{\p\Psi}, \\
X_5= F_3(t)\frac{\p}{\p y}-\frac{\epsilon\,x}{Re}\Big(\alpha
F_2(t)+(Re-2\alpha t)F_2'(t)\Big)\frac{\p}{\p\Psi} \ea\] and \be
\label {3} X_6=\ln \Big(t-\frac{Re}{2\alpha}\Big)\Big(y\frac{\p}{\p
x}-x\frac{\p}{\p y}\Big)-\frac{\alpha\epsilon}{Re}(x^2+y^2)\frac{\p}{\p\Psi}.
\ee Obviously the Lie symmetries $X_i, \ i=1,\dots,5$ coincide (up
to notations) with those in \cite{mandal-2025} (there are misprints in (21)). However, the Lie
symmetry (\ref{3}) cannot be derived from the Lie symmetries listed
in (21)\cite{mandal-2025}.

 In the
next step, the authors analyse two-dimensional PDE
(23)\cite{mandal-2025}, which is nothing else but  Eq.(\ref{1}) in
the stationary case: \be \label {2} \ba \frac{Re}{\epsilon^2}\Big(
\frac{\partial f}{\partial y} \frac{\partial^{3}f}{\partial
x\partial y^{2}}-\frac{\partial f}{\partial x}
\frac{\partial^{3}f}{\partial y^{3}} \Big)
-\frac{Re}{\epsilon^2}\Big( \frac{\partial f}{\partial x}
\frac{\partial^{3}f}{\partial y\partial x^{2}}-\frac{\partial
f}{\partial y} \frac{\partial^{3}f}{\partial x^{3}} \Big)=\\
-\frac{1}{D_A}\Big( \frac{\partial^{2}f}{\partial x^{2}}+
\frac{\partial^{2}f}{\partial y^{2}} \Big)+\lambda
\Big(\frac{\partial^{4}f}{\partial
x^{4}}+\frac{\partial^{4}f}{\partial
y^{4}} +2\frac{\partial^{4}f}{\partial x^2\partial y^{2}}\Big)\\
+\frac{\alpha}{\epsilon}\Big( 2(\frac{\partial^{2}f}{\partial
x^{2}}+ \frac{\partial^{2}f}{\partial
y^{2}})+x(\frac{\partial^{3}f}{\partial
x^{3}}+\frac{\partial^{3}f}{\partial x \p
y^{2}})+y(\frac{\partial^{3}f}{\partial
y^{3}}+\frac{\partial^{3}f}{\partial y \p x^{2}}) \Big) \ea
 \ee In Theorem 2
\cite{mandal-2025}, the authors claim that the generator
$Y=S(y)\frac{\p}{\p f} $ with the  function $S$ satisfying a fourth-order ODE  is  the
only nonclassical symmetry  of PDE (\ref{2}). Obviously, this
statement is incorrect. In fact, taking into account that  PDE
(\ref{2}) is symmetric with respect to the variables $x$ and $y$,
one immediately concludes that  $X=S(x)\frac{\p}{\p f} $ is a
nonclassical symmetry as well. In reality, the authors used an
incorrect definition of nonclassical symmetry of PDEs. Before
formulation of a rigorous definition, it should be noted that each
nonclassical symmetry is defined up to an arbitrary multiplier (see
the proof in Section 3.1 of \cite{ch-se-pl-book}). It means that the
generator $M(x,y,f)Q$ (here $M$ is an arbitrary smooth function) is
a nonclassical symmetry of PDE (\ref{2}) provided the generator $Q$
is such a symmetry. On the other hand, it is obvious that
$Q=\frac{\p}{\p f} $ is a Lie symmetry of PDE (\ref{2}), therefore
that is automatically a nonclassical symmetry. Now one concludes
that each  generator  of the form   $M(x,y,f)\frac{\p}{\p f}$  (not
only $S(y)\frac{\p}{\p f} $!) is a nonclassical symmetry of  PDE
(\ref{2}). However, all these symmetries are equivalent to the Lie
symmetry  $Q=\frac{\p}{\p f} $.

In order to find nonclassical symmetries (not only  those that are
equivalent to Lie symmetries), one needs to use the correct
definition for an arbitrary k-th order PDE \be\label{lb37}
L\lf(t,x,u,\underset{1}u,\dots,\underset{k}u\rg)=0, \ \ k\geq1,\ee
 where  $u=u(x,y)$ is an unknown function,
 $\mbox{\raisebox{-1.2ex}{$\stackrel{\displaystyle  
u}{\scriptstyle s}$}}$ means a totality of $s$-order derivatives of
$u(x,y)$ ($s=1,2,\dots,k$) and $L$ is a given  smooth function.


\begin{definition}\cite[Section 3.1]{ch-se-pl-book}
 Operator
\be\label{lb5}
Q=\xi^1(x,y,u)\p_x+\xi^2(x,y,u)\p_{y}+\eta(x,y,u)\p_u,
\ee where  $\xi^1(x,y,u), \xi^2(x,y,u)$ and $\eta(x,y,u)$ are given
smooth functions,
  is called $Q$-conditional (nonclassical) symmetry
of PDE   (\ref{lb37}) if the following invariance criteria  is
satisfied:
\be\label{lb13} \underset{k}Q (L) \,  
 \Big\vert_{\cal{M}}=0,\ee where the differential operator  $\underset{k}Q$
 is the   $k$-order prolongation of  operator
(\ref{lb5})  and the manifold\index{manifold}
 ${\cal{M}}$ is defined by the system of equations
  \[ L=0, \quad Q(u)=0,  \quad \frac{\p^{p+q}Q(u)}{\p\, x^py^q}=0,  \quad 1\leq p+q\leq k-1\]
  in the prolonged space of the variables \[x, \ y, \ u, \ \underset{1}u, \ \dots, \ \underset{k}u.\]
\end{definition}
The main peculiarity of the definition consists in differential
consequences of the equation
 \[Q(u)\equiv \xi^1u_x+\xi^2u_y - \eta = 0, \]
  which must be taking into
account. Note that the same definition is formulated in words in the
book \cite{bl-anco-10} (see Section 5.2.2 therein), which is cited
in \cite{mandal-2025}. In the case of PDE (\ref{2}),
all differential consequences
\[\frac{\p^{p+q}Q(f)}{\p\, x^py^q}=0,  \quad 1\leq p+q\leq 3\]
must be taking into account. It was not done in \cite{mandal-2025},
therefore the result obtained therein is trivial.

Finally, it should be highlighted that special case, $\xi^2=0, \
\xi^1=1$, which is separately examined in \cite{mandal-2025}, is
known as 'no-go case'. It is well-known that this case always leads
to the system of determining equations, which consists of {\it a
single PDE}, and this contradicts to the system of equations
presented on P.6  in \cite{mandal-2025}. Moreover, the single
determining PDE
 is related   to the initial equation. In the case of an arbitrary
 evolution equation,  the
 corresponding determining equation is reducible to the given
 equation by a chain of substitutions (see the proof in \cite{zhdanov}). As a result, one can claim
 that the search for nonclassical symmetries in no-go case is
 equivalent to solving the given equation. Notably, some progress in solving this problem
 was achieved  in the case of {\it systems} of PDEs
 \cite{cherniha21b,ch-da-AAM-2023}.

\end{document}